\documentclass[journal]{IEEEtran}

\usepackage{cite}      

\usepackage{graphicx}  

\usepackage{amsmath}   
\usepackage{amssymb}
\usepackage{dsfont}
\usepackage{subfigure}
\newtheorem{theorem}{Theorem}

\newtheorem{assumption}{Assumption}
\allowdisplaybreaks
\begin{document}
\title{Recoverability Analysis for Modified Compressive Sensing with
Partially Known Support}

\author{Jun~Zhang, Yuanqing~Li, Zhu~Liang~Yu and
Zhenghui~Gu
\thanks{J.~ZHANG, Y.~LI, Z.~l.~YU and Z.~GU are with the College of Automation Science and
Engineering, South China University of Technology, Guangzhou,
510640, China. J.~ZHANG is also with College of Information
Engineering, Guangdong University of Technology, Guangzhou, 510006, China, e-mail: jzhang@gdut.edu.cn.}
\thanks{}
}
%
%
%

\maketitle

\begin{abstract}
The recently proposed modified-compressive sensing (modified-CS), which utilizes the partially known support as
prior knowledge, significantly improves the performance of recovering sparse signals. However, modified-CS depends heavily on the reliability of the known support. An important problem, which must be studied further, is the recoverability of modified-CS when the known support contains a number of errors. In this letter, we analyze the
recoverability of modified-CS in a stochastic framework. A sufficient and necessary condition is established for exact recovery of a sparse signal. Utilizing this condition, the recovery probability that reflects the recoverability of modified-CS can be computed explicitly for a sparse signal with $\ell$ nonzero entries, even though the known support exists some errors. Simulation experiments have been carried out to validate our theoretical results.
\end{abstract}


\section{Introduction}
Compressive Sensing (CS) allows exact recovery of a sparse signal using only a limited number of random measurements. A central problem in CS is the following: given an $m \times n$ matrix $\textbf{A}$ ($m < n$), and a measurement vector $\textbf{y}=\textbf{A}\textbf{x}^*$, recover $\textbf{x}^{*}$. To deal with this problem, the most extensively studied recovery
method is the $\ell_1$-minimization approach (Basis Pursuit) \cite{Do_CS, Ca_CS2, Ca_CS, Do_BP, candes2005decoding}
\begin{equation}\label{p1}
 \mathop {\min }\limits_\textbf{x} {\left\| {{\textbf{x}}} \right\|_1}\quad \quad s.t\quad \textbf{y} =
\textbf{Ax}
\end{equation}
This convex problem can be solved efficiently; moreover, $\mathcal{O}(\ell\log (n/\ell))$ probabilistic measurements are sufficient for it to recover a $\ell$-sparse vector $\textbf{x}^{*}$ (i.e., all but at most $\ell$ entries are zero) exactly.

Recently, Vaswani and Lu \cite{Va_MCS, Va_MCS_C, Va_MCS_C2,
Va_MCS_C3}, Miosso \cite{CM_IRLS, von2007compressed}, Wang and Yin \cite{Wa_CSISD,
Wa_ECS}, Friedlander et.al \cite{friedlander2010recovering}, Jacques \cite{jacques2010short} have shown that exact recovery based on fewer measurements
than those needed for the $\ell_1$-minimization approach
is possible when the support of $\textbf{x}^{*}$ is partially known. The
recovery is implemented by solving the optimization
problem.
\begin{equation} \label{eq:2}
\mathop {\min }\limits_\textbf{x} {\left\| {{\textbf{x}^{
}_{\textbf{T}^c}}} \right\|_1}\quad \quad s.t\quad \textbf{y} =
\textbf{Ax}
\end{equation}
where \textbf{T} denotes the "known" part of support, $\textbf{T}^{c}=[1, ..., n]\setminus \textbf{T}$,
$\textbf{x}^{ }_{\textbf{T}^c}$ is a column vector composed of the
entries of $\textbf{x}$ with their indices being in $\textbf{T}^c$. This method is named modified-CS \cite{Va_MCS} or
truncated $\ell_1$ minimization \cite{Wa_CSISD}. One application of the
modified-CS is the recovery of (time) sequences of sparse signals,
such as dynamic magnetic  resonance imaging (MRI) \cite{Va_MCS_C2, Va_MCS_C3}. Since the support evolve slowly over time, the previously
recovered support can be used as known part for later
reconstruction.

As an important performance index of modified-CS, its
recoverability, i.e., when is the solution of (\ref{eq:2}) equal to $\textbf{x}^{*}$, has been discussed in several papers. In \cite{Va_MCS}, a sufficient condition on the recoverability was obtained based on restricted isometry property. From the view of \emph{t}-null space property, another sufficient condition to recover $\ell$-sparse vectors was proposed in \cite{Wa_CSISD}. However, there always exist some signals that do not satisfy these conditions but still can be recovered. Specifically, in real-world applications, the known support often contains some errors. The existing sufficient conditions can not reflect accurately the recoverability of modified-CS in many cases. Therefore, it is necessary to develop alternative techniques for analyzing the recoverability of modified-CS.

In this letter, a sufficient and necessary condition (SNC) on the recoverability of modified-CS is derived. Then, we discuss the recoverability of modified-CS in a probabilistic way. The main advantage of our work is that, for a randomly given vector $\textbf{x}^{*}$ with $\ell$ nonzero entries, the exact recovery percentage of modified-CS can be computed explicitly under a given matrix $\textbf{A}$ and a randomly given $\textbf{T}$ that satisfied $|\textbf{T}|=p$ but includes $p_1$ errors, where $|\textbf{T}|$ denotes the size of the known support $\textbf{T}$. Hence, this paper provides a quantitative index to measure the reliability of modified-CS in real-world applications. Simulation experiments validate our results.

\section{Probability Estimation On Recoverability of Modified-CS}
In this section, a SNC
on the recoverability of modified-CS is derived. Based on this condition, we discuss the estimation of the probability that the vector $\textbf{x}^{*}$ can be recovered by modified-CS. We name this probability as recovery probability.

\subsection{A Sufficient and Necessary Condition for Exact Recovery}

Firstly, some notations are given in the follows. The
support of $\textbf{x}^*=(x_1^{*}, ...,
x_n^{*})^{T}$ is denoted by $\textbf{N}$, i.e.
$\textbf{N} \buildrel \Delta \over =\{j|x_j^{*}\neq 0\}$. Suppose
$\textbf{N}$ can be split as
$\textbf{N}=\textbf{T}\cup\boldsymbol{\Delta}\setminus\boldsymbol{\Delta_e}$,
where $\boldsymbol{\Delta}\buildrel \Delta \over = \textbf{N} \setminus \textbf{T}$ is the
unknown part of the support and $\boldsymbol{\Delta_e} \buildrel \Delta \over =
\textbf{T} \setminus \textbf{N}$ is set of errors in the known part support $\textbf{T}$.
The set operations $\cup$ and $\setminus$ stand for set union and
set except respectively.

Let $\textbf{x}^{(1)}$ denote the solution of the model
(\ref{eq:2}) and $\textbf{F}$ denote the set of all subsets of
$\boldsymbol{\Delta}$. A SNC is given in the following theorem,
which is an extension of a result in \cite{li2006probability}.

\begin{theorem}
\label{th1}
 For a given vector $\textbf{x}^*$, $\textbf{x}^{(1)}=\textbf{x}^*$, if and only if $ \forall \textbf{I} \in \textbf{F}$, the
optimal value of the objective function of the following
optimization problem is greater than zero, provided that this
optimization problem is solvable:
\begin{equation} \label{eq:3}
\begin{array}{l}
\begin{split}
 \mathop {\min }\limits_{\boldsymbol{\delta}} &\sum\limits_{k \in ({\textbf{T}^c}\backslash \textbf{I})} {\left| {{{\delta} _k}}
\right| - \sum\limits_{k \in \textbf{I}} {\left| {{{\delta} _k}} \right|} },  \,\,\, s.t.\\
 &\textbf{A}\boldsymbol{\delta}  = \textbf{0},\;{\left\| \boldsymbol{\delta}  \right\|_1} = 1 \\
 &{{\delta} _k}{x}_k^ *  > 0\;\;for\;k \in \textbf{I} \\
 &{{\delta} _k}{x}_k^ *  \le 0\;\;for\;k \in \boldsymbol{\Delta} \backslash \textbf{I} \\
 \end{split}
\end{array}
\end{equation}
where $\boldsymbol{\delta}=(\delta_1, ..., \delta_n)^{T} \in
\textbf{R}^n$.
\end{theorem}

The proof of this theorem is given in Appendix I.

{\textbf{Remark 1: }For a given measurement matrix $\textbf{A}$,
the recoverability of the sparse vector $\textbf{x}^*$ based on the
model (\ref{eq:2}) depends only on the index set of nonzeros of
$\textbf{x}^*$ in $\textbf{T}^c$ and the signs of these nonzeros. In
other words, the recoverability relies only on the sign pattern of
$\textbf{x}^*$ in $\textbf{T}^c$ instead of the magnitudes of these
nonzeros.

{\textbf{Remark 2: }It follows from the proof of Theorem 1 that, even if
$\textbf{T}$ contains several errors, Theorem \ref{th1} still holds.


\subsection{ Probability Estimation for Recoverability of the Modified-CS}

In this subsection, we utilize Theorem 1 to estimate the recovery probability, i.e., the conditional probability
$\textbf{P}(\textbf{x}^{(1)}=\textbf{x}^{*};
\|\textbf{x}^{*}\|_{0}=\ell, |\textbf{T}|=p,
|\boldsymbol{\Delta}_e|=p_{1}, \textbf{A})$, where $\|\textbf{x}^{*}\|_0$ is defined as the number of nonzero entries
of $\textbf{x}^{*}$, $|\textbf{T}|$ and $|\boldsymbol{\Delta}_e|$ denote the size of $\textbf{T}$ and $\boldsymbol{\Delta}_e$ respectively. Let $\textbf{G}$ denote the index set $\{1, 2, ..., n\}$, it is easy to know that there are $C_n^\ell( =
\frac{{n!}}{{\ell!(n - \ell)!}})$ index subsets of $\textbf{G}$ with
size $\ell$. We denote these subsets as
$\textbf{G}_j^{(\ell)}$, $j=1, ..., C_n^\ell$. For each
$\textbf{G}_j^{(\ell)}$, there are
$C_\ell^{p_2}$
subsets with size $p_2=(p-p_{1})$. We denote these subsets as
$\textbf{N}_s^{(p_{2})}, \; s = 1, ..., C_\ell^{p_2}$. At the same time, for the set $\textbf{G}\backslash
\textbf{N}$ (the index set of the zero entries of $\textbf{x}^{*}$),
there are $C_{n-\ell}^{p_1}$ subsets with size $p_1$. These
subsets are denoted as $\textbf{H}_i^{(p_{1})}, \; i = 1, ...,
C_{n-\ell}^{p_1}$. Without loss of
generality, we have the following assumption.

\begin{assumption}
The index set $\textbf{N}$ of the $\ell$ nonzero entries of
$\textbf{x}^*$ can be one of the $C_n^\ell$ index sets
$\textbf{G}_j^{(\ell)}$, $j=1, ..., C_n^\ell$, with equal
probability. The index set $\boldsymbol{\Delta}_e$ of $p_1$ errors
in known support can be one of the $C_{n-\ell}^{p_1}$ index sets
$\textbf{H}_i^{(p_{1})}$, $i = 1, ..., C_{n-\ell}^{p_1}$, with equal
probability. The index set $\textbf{T} \backslash
\boldsymbol{\Delta}_e$ of $p_2$ nonzero entries can be one of the
$C_{\ell}^{p_2}$ index sets $\textbf{N}_s^{(p_{2})}$, $s = 1, ...,
C_{\ell}^{p_2}$, with equal probability. All the nonzero entries of the vector $\textbf{x}^*$ take either
positive or negative sign with equal probability.
\end{assumption}

For a given vector $\textbf{x}^*$ and the known
support $\textbf{T}$, there is a sign column vector
$\textbf{t}=sign(\textbf{x}_{\textbf{T}^{c}}^{*})\in
\textbf{R}^{n-p}$ in $\textbf{T}^c$. The recoverability of the
vector $\textbf{x}^*$ only relates with the sign column vector
$\textbf{t}$ (see Remark 1). Under the conditions that the index set
of the nonzero entries of $\textbf{x}^*$ is $\textbf{G}_j^{(\ell)}$
and the known support is $\textbf{N}_s^{(p_2)} \cup
\textbf{H}_i^{(p_1)}$, then there are $2^{\ell-p_2}$ sign column
vectors. Among these sign column vectors, suppose that $w_{s,i}^j$
sign column vectors can be recovered, then
$\frac{{w_{s,i}^j}}{{{2^{\ell - p_2}}}}$ is the probability of the
vector $\textbf{x}^*$ being recovered by solving the modified-CS.
Hence, following Assumption 1, the recovery probability
is calculated by
\begin{equation}
\label{eq888}
\begin{split}
{\bf{P}}({{\bf{x}}^{(1)}} = {{\bf{x}}^*}; &{\left\| {{{\bf{x}}^*}}
\right\|_0} = \ell ,|{\bf{T}}| = p,\left| {{\boldsymbol{\Delta}
_e}} \right| = {p_1},{\bf{A}}) \\
&= \sum\limits_{j = 1}^{{\bf{C}}_n^\ell } {\frac{1}{{{\bf{C}}_n^\ell
}}\sum\limits_{s = 1}^{{\bf{C}}_\ell ^{{p_2}}}
{\frac{1}{{{\bf{C}}_\ell ^{{p_2}}}}\sum\limits_{i =
1}^{\textbf{C}_{n - \ell }^{{p_1}}} {\frac{1}{{\textbf{C}_{n - \ell
}^{{p_1}}}}\frac{{w_{s,i}^j}}{{{2^{\ell - {p_2}}}}}} } }
\end{split}
\end{equation}
where $\ell=1, ..., m$, $p=0, ..., \ell$, $p_{1}=0, ..., p$ and $p_{2}=p-p_{1}$.

Because the measurement matrix $\textbf{A}$ is known, we can
determine $w_{s,i}^{j}$ in (\ref{eq888}) by checking whether the
SNC (\ref{eq:3}) is satisfied for all
the $2^{\ell-p_{2}}$ sign column vectors corresponding to the index
set $\textbf{G}_j^{(\ell)}$, $\textbf{H}_i^{(p_{1})}$ and
$\textbf{N}_s^{(p_2)}$. Now we present a simulation example to demonstrate the validity of the
probability estimation by (\ref{eq888}) through comparing it with simulation results.

\emph{Example 1: }
Suppose $\textbf{A} \in \textbf{R}^{7 \times 9}$ was taken according to the
uniform distribution in [-0.5, 0.5]. All nonzero entries of the
sparse vector $\textbf{x}^*$ were drawn from a uniform distribution
valued in the range [-1, +1]. Without loss of generality, we set $p=2$. For a vector
$\textbf{x}^{*}$ with $\ell$ nonzero entries, where $\ell$=2, 3,
..., 7, we calculated the recovery probabilities by (\ref{eq888}),
where $p_{1}=0, 1, 2$ respectively. For every $\ell$
($\ell = 2, ..., 7 $) nonzero entries, we also sampled 1000 vectors
with random indices. For each vector, we solved the modified-CS with
a randomly given $\textbf{T}$, whose size equals to $p$ but contains
$p_1$ errors, and checked whether the solution is equal to the true
vector. Suppose that
$n_p^\ell$ vectors can be recovered, we calculated the ratio
$p_p^\ell = \frac{{n_p^\ell }}{{1000}}$ as the recovery probability $\boldsymbol{\hat
P}(\textbf{x}^{(1)}=\textbf{x}^{*}; \|\textbf{x}^{*}\|_{0}=\ell,
|\textbf{T}|=p, |\boldsymbol{\Delta}_e|=p_{1}, \textbf{A})$. The experimental results
are presented in Fig. \ref{fig1}. Therein, solid curves denote the theoretic
recovery probability estimated by (\ref{eq888}). Dotted curves
denote probabilities $\boldsymbol{\hat
P}(\textbf{x}^{(1)}=\textbf{x}^{*}; \|\textbf{x}^{*}\|_{0}=\ell,
|\textbf{T}|=p, |\boldsymbol{\Delta}_e|=p_{1}, \textbf{A})$.
Experimental results show that the theoretical estimates
fit the simulated values very well.
\begin{figure}
\centering
\includegraphics[width=3.5 in]{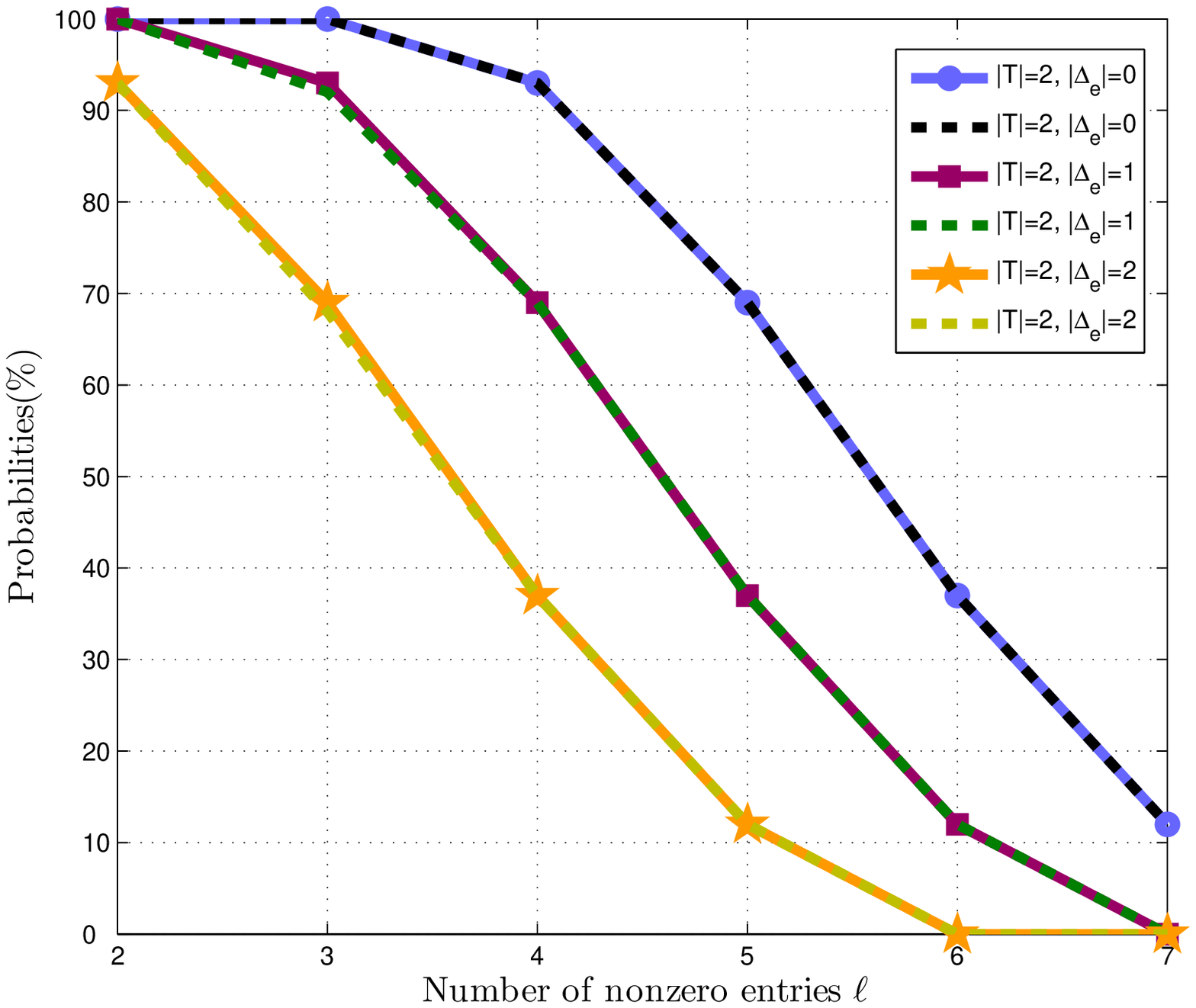}
\caption{Comparison of theoretical and simulation results on
recovery probability. Solid curves: probability curves of
$\textbf{P}(\textbf{x}^{(1)}=\textbf{x}^{*};
\|\textbf{x}^{*}\|_{0}=\ell, |\textbf{T}|=p,
|\boldsymbol{\Delta}_e|=p_{1}, \textbf{A})$ obtained by
(\ref{eq888}); dotted curves: probability curves obtained by a
random sampling. The three pairs of solid and dotted curves from the
top to the bottom correspond to $|\textbf{T}| = 2$, $ \left|
{{\boldsymbol{\Delta} _e}} \right| = 0$, 1, 2
respectively.}\label{fig1}
\end{figure}

However, the computational burden to calculate (\ref{eq888}) increases exponentially as the problem dimensions increase. As mentioned above, for each sign column vector and the corresponding index
sets, we denote the quads $[\textbf{G}_j^{(\ell)}, \textbf{N}_s^{(p_2)}, \textbf{H}_i^{(p_{1})}, \textbf{t}_\tau]$, where $j=1,...,C_n^\ell$, $s=1,...,C_{\ell}^{p_2}$, $i=1,...,C_{n-\ell}^{p_1}$ and $\tau = 1,...,2^{\ell-p_2}$. Suppose $\textbf{\textit{Z}}$ is a set composed by all the quads, there are $C_n^\ell C_{\ell}^{p_2} C_{n-\ell}^{p_1}2^{\ell-p_2}$ elements in $\textbf{Z}$. For each element of $\textbf{Z}$, if the sign column vector $\textbf{t}_\tau$ can be recovered by the modified-CS with a given matrix $\textbf{A}$ and a known support $\textbf{T} = \textbf{N}_s^{(p_2)} \cup
\textbf{H}_i^{(p_1)}$, we call the quad can be recovered. In (\ref{eq888}), the estimation of recover probability need to check the total number of quads in $\textbf{Z}$. Hence, when $n$ increases, the computational burden will increase exponentially. To avoid this problem, we state the following Theorem.
\begin{theorem}
\label{th2}
Suppose that $M$ quads are randomly taken from set $\textbf{Z}$, where $M$ is a large positive integer ($M \ll C_n^\ell C_{\ell}^{p_2}C_{n-\ell}^{p_1}2^{\ell-p_2}$), and $K$ of the $M$ quads can be recovered by solving modified-CS. Then
\begin{equation}\label{app}
{\bf{P}}({{\bf{x}}^{(1)}} = {{\bf{x}}^*}; {\left\| {{{\bf{x}}^*}}
\right\|_0} = \ell ,|{\bf{T}}| = p,\left| {{\boldsymbol{\Delta}
_e}} \right| = {p_1},{\bf{A}}) \backsimeq \frac{K}{M}
\end{equation}
\end{theorem}

The proof of this theorem is given in Appendix II.

{\textbf{Remark 3:} In real-world applications, by sampling randomly $M$ sign vectors with $\ell$ nonzero entries, we can check the number of the vectors that can be exactly recovered by modified-CS with a random known support $\textbf{T}$ whose size is $p$ but contains $p_1$ errors. Suppose $K$ sign vectors can be recovered, the recovery probability $\textbf{P}(\textbf{x}^{(1)}=\textbf{x}^{*};
\|\textbf{x}^{*}\|_{0}=\ell, |\textbf{T}|=p,
|\boldsymbol{\Delta}_e|=p_{1}, \textbf{A})$ can be computed approximately through calculating the ratio of $K/M$.

From the proof of Theorem 2, the number of samples $M$, which controls the precision in the approximation  of (\ref{app}), is related to the two-point distribution of $\upsilon_k$ other than the size of $\textbf{Z}$. Thus, there is no need for $M$ increasing exponentially as $n$ increases. In the following example 2, this conclusion as well as the conclusion in Theorem (\ref{th2}) are demonstrated.

\emph{Example 2: } In this example, according to the uniform distribution in [-0.5, 0.5], we randomly generate three matrices $\textbf{A}_i \in R^{m \times n}$ ($i=1,2,3$) with ($m$, $n$)=(7, 9), (52, 128) and (181, 1280) respectively. For matrices $\textbf{A}_1$, $\textbf{A}_2$ and $\textbf{A}_3$, we set ($\ell$, $p$, $p_1$)=(4, 2, 1), (20, 8, 3) and (60, 32, 4) respectively. As $n$ increases in the three cases, the number of sign vectors increases exponentially. For example, for $(m, n, \ell, p, p_1)=(7, 9, 4, 2, 1)$ and $(52, 128, 20, 8, 3)$, the set $\textbf{Z}$ contains approximately $2.02 \times 10^4$ and $1.24 \times 10^{37}$ elements respectively. Hence, for the three cases, we estimate the probabilities $\textbf{P}(\textbf{x}^{(1)}=\textbf{x}^{*};\|\textbf{x}^{*}\|_{0}=\ell, |\textbf{T}|=p,|\boldsymbol{\Delta}_e|=p_{1}, \textbf{A})$ by (\ref{app}). For each case, we sample $M$=100, 500, 1000, 5000, 10000 respectively. The resultant probability estimates depicted in Fig. \ref{fig2} indicate that 1) the estimation precision of (\ref{app}) is stable in our experiments with different number of samples. Therefore, we may just need very few samples to obtain the satisfied estimation precision in real-world applications; 2) as $n$ increases in three cases, the number of samples $M$ don't need an exponential increase.

\begin{figure}
\centering
\includegraphics[width=3.5 in]{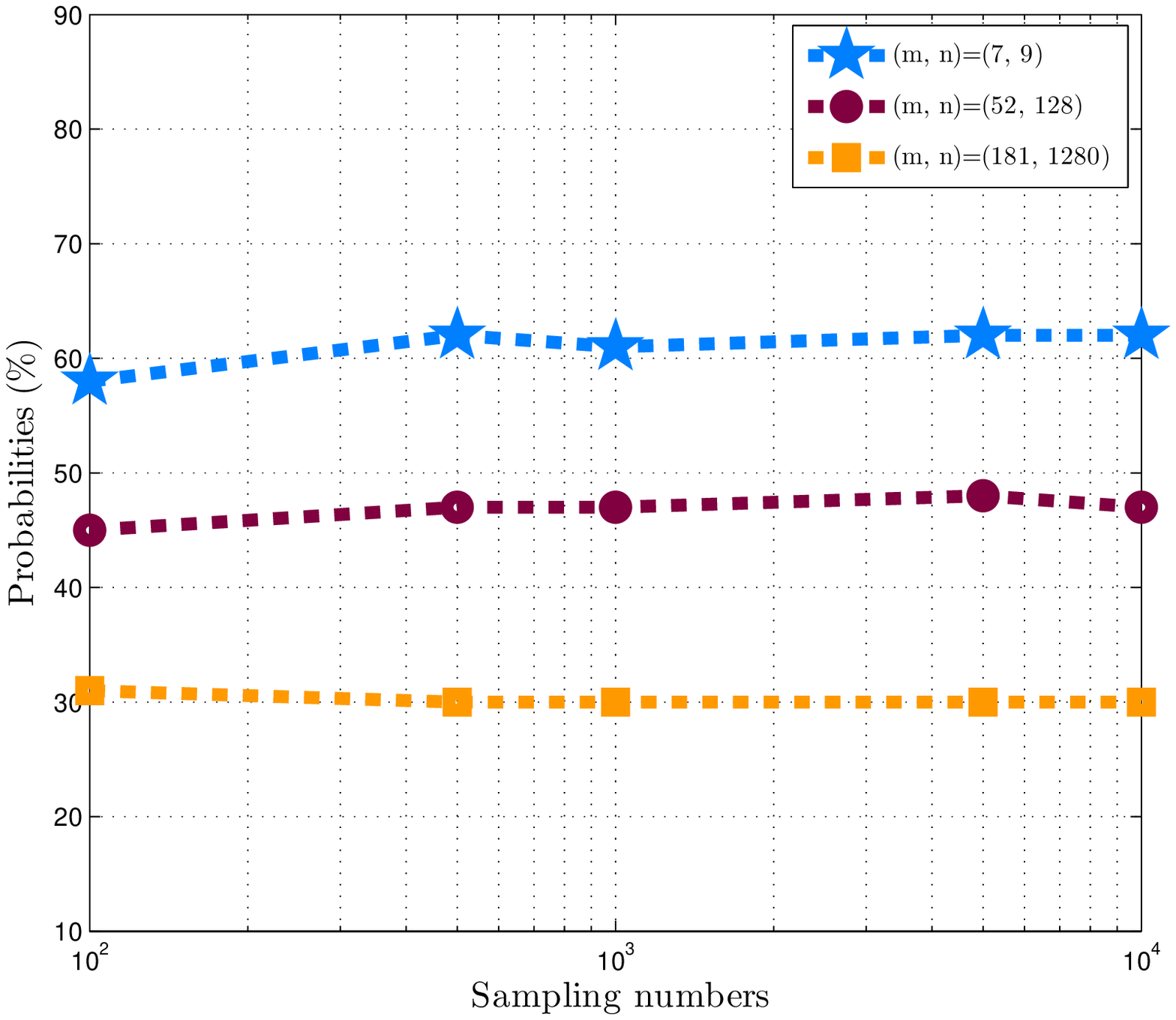}
\caption{Probabilities curves obtained in example 2. The horizontal axis represents the sampling numbers. The vertical axis represents the probabilities $\textbf{P}(\textbf{x}^{(1)}=\textbf{x}^{*};
\|\textbf{x}^{*}\|_{0}=\ell, |\textbf{T}|=p,
|\boldsymbol{\Delta}_e|=p_{1}, \textbf{A})$ obtained by
(\ref{app}).The three curves from the
top to the bottom correspond to $(m, n, \ell, p, p_1)=(7, 9, 4, 2, 1)$, $(52, 128, 20, 8, 3)$ and $(181, 1280, 60, 32, 4)$ respectively.}\label{fig2}
\end{figure}

\section{Conclusion}
In this letter we study the recoverability of the
modified-CS in a stochastic framework. A sufficient and necessary
condition on the recoverability is presented. Based on this
condition, the recovery
probability of the modified-CS can be estimated explicitly. It is worth mentioning that Theorem 1 can be easy to extend to the weighted-$\ell_1$ minimization approach that was proposed in \cite{khajehnejad2010analyzing} for nonuniform sparse model. Moreover, the recovery probability estimation provides alternative way to find (numerically) the optimal set of weights in the weighted-$\ell_1$ minimization approach, which has the largest recovery probability to recover the signals.


\appendices
\section{Proof of Theorem 1}
\begin{proof}
\textbf{Necessity:} Suppose that $\textbf{x}^{(1)}=\textbf{x}^*$.
Thus $\textbf{x}^*$ is the optimal solution and
$\|\textbf{x}_{\textbf{T}^c}^*\|_{1}$ is the optimal value of the
optimization problem in (\ref{eq:2}).

For a subset $\textbf{I} \in \textbf{F}$, when (\ref{eq:3}) is
solvable, there is at least a feasible solution. For a feasible
solution $\boldsymbol{\delta}$ of (\ref{eq:3}), it can be proved
that $\textbf{x}^{*}+t\boldsymbol{\delta}$ is a solution of the
constraint equation of (\ref{eq:2}), where $t$ is a constant. In the
following, we suppose $t<0$ with sufficiently small absolute value.
Then we have
\begin{equation}
\label{al1}
\begin{split}
{\left\| {\textbf{x}_{{\textbf{T}^c}}^ *  + t{\boldsymbol{\delta}
_{{\textbf{T}^c}}}} \right\|_1} &= \sum\limits_{k \in \textbf{I}}
{\left| {{x}_k^ *  + t{{\delta}_k}} \right| + \sum\limits_{k \in
\boldsymbol{\Delta} \backslash \textbf{I}} {\left| {{x}_k^
*  + t{{\delta} _k}} \right|}} + \\
&\quad\sum\limits_{k \in {T^{c} \backslash \boldsymbol{\Delta}}}
{\left|
{t{{\delta} _k}} \right|}\\
&=\sum\limits_{k \in \textbf{I}} {\left| {{x}_k^ * } \right|} -
\left| t \right|\sum\limits_{k \in \textbf{I}} {\left| {{{\delta}
_k}} \right|} + \sum\limits_{k \in \boldsymbol{\Delta} \backslash
\textbf{I}} {\left| {{x}_k^ * } \right|} +\\
&\quad\left| t \right|\sum\limits_{k \in \boldsymbol{\Delta}
\backslash \textbf{I}} {\left| {{{\delta} _k}} \right| +
\left| t \right|} \sum\limits_{k \in {T^{c} \backslash \boldsymbol{\Delta}}} {\left| {{{\delta} _k}} \right|}\\
&= {\left\| {\textbf{x}_{{\textbf{T}^c}}^ * } \right\|_1} + \left| t
\right|(\sum\limits_{k \in ({\textbf{T}^c}\backslash \textbf{I})}
{\left| {{{\delta} _k}} \right| - \sum\limits_{k \in \textbf{I}}
{\left| {{{\delta} _k}} \right|} } )
\end{split}
\end{equation}

Since $\textbf{x}^*$ is the optimal solution of the optimization
problem (\ref{eq:2}), it follows from (\ref{al1}) that

\begin{equation}
{\left\| {\textbf{x}_{{\textbf{T}^c}}^ * } \right\|_1} + \left| t
\right|(\sum\limits_{k \in ({\textbf{T}^c}\backslash \textbf{I})}
{\left| {{{\delta} _k}} \right| - \sum\limits_{k \in \textbf{I}}
{\left| {{{\delta} _k}} \right|} } )> {\left\|
{\textbf{x}_{{\textbf{T}^c}}^ * } \right\|_1}
\end{equation}

Thus,
\begin{equation}
\sum\limits_{k \in ({\textbf{T}^c}\backslash \textbf{I})} {\left|
{{{\delta} _k}} \right| - \sum\limits_{k \in \textbf{I}} {\left|
{{{\delta} _k}} \right|} }  > 0
\end{equation}
The necessity is proved.

 \textbf{Sufficiency:} Suppose that
$\textbf{x}^{\dag}$ is a solution of the constraint equation in
(\ref{eq:2}), which is different from $\textbf{x}^{*}$. Then
$\textbf{x}^{\dag}$ can be rewritten as
\begin{equation}
\label{al2} \textbf{x}^{\dag} = {\textbf{x}^ * } + {t^ *
}\boldsymbol{\delta},
\end{equation}
where $\boldsymbol{\delta}  = \frac{{({\textbf{x}^ * } -
\textbf{x}^{\dag})}}{{{{\left\| {{\textbf{x}^ * } -
\textbf{x}^{\dag}} \right\|}_1}}}$, ${t^ * } =  - {\left\|
{{\textbf{x}^ * } - \textbf{x}^{\dag}} \right\|_1}\not=0$.

 Now we define
an index set $\textbf{I}$,
\begin{equation} \label{setI}
\textbf{I} = \{ k|k \in \boldsymbol{\Delta},\; sign({x}_k^ * ) =
sign({{\delta} _k})\}.
\end{equation}

From (\ref{al2}), we have
\begin{equation}
\label{eq13}
\begin{split}
{\left\| {\textbf{x}_{{\textbf{T}^c}}^\dag } \right\|_1}&={\left\|
{\textbf{x}_{{\textbf{T}^c}}^ *  + t^{*}{\boldsymbol{\delta}
_{{\textbf{T}^c}}}} \right\|_1} \\
&= \sum\limits_{k \in \textbf{I}} {\left| {{x}_k^ *  +
t^{*}{{\delta} _k}} \right|} + \sum\limits_{k \in
\boldsymbol{\Delta} \backslash \textbf{I}} {\left| {{x}_k^ {*}
+t^{*}{{\delta} _k}} \right|} +\\
&\quad \sum\limits_{k \in {\textbf{T}^{c}\backslash
\boldsymbol{\Delta}}} {\left| {t^{*}{{\delta}
_k}} \right|} \\
&\geq\sum\limits_{k \in \textbf{I}} {\left| {{x}_k^ * } \right|}  -
\left| t^{*} \right|\sum\limits_{k \in \textbf{I}} {\left|
{{{\delta} _k}} \right|}  + \sum\limits_{k \in \boldsymbol{\Delta}
\backslash \textbf{I}} {\left| {{x}_k^ * } \right|} +\\
&\quad\quad\quad\quad\left| t^{*} \right|\sum\limits_{k \in
\boldsymbol{\Delta} \backslash \textbf{I}} {\left|
{{{\delta} _k}} \right| + \left| t^{*} \right|} \sum\limits_{k \in {\textbf{T}^{c}\backslash \boldsymbol{\Delta}}} {\left| {{{\delta _k}}} \right|}\\
&= {\left\| {\textbf{x}_{{\textbf{T}^c}}^ * } \right\|_1} + \left|
t^{*} \right|(\sum\limits_{k \in ({\textbf{T}^c}\backslash
\textbf{I})} {\left| {{{\delta} _k}} \right| - \sum\limits_{k \in
\textbf{I}} {\left| {{{\delta} _k}} \right|} } )
\end{split}
\end{equation}

It can be easily proved that for the defined index set $\textbf{I}$
in (\ref{setI}), $\textbf{I} \in \textbf{F}$ and
$\boldsymbol{\delta}$ is a feasible solution of (\ref{eq:3}). From
the condition of the theorem, we have
\begin{equation} \label{eq200}
\sum\limits_{k \in ({\textbf{T}^c}\backslash \textbf{I})} {\left|
{{{\delta} _k}} \right| - \sum\limits_{k \in \textbf{I}} {\left|
{{{\delta} _k}} \right|} }  > 0.
\end{equation}

Combining (\ref{eq13}) and (\ref{eq200}), we have
\begin{equation}
\label{eq142} {\left\| {\textbf{x}_{{\textbf{T}^c}}^\dag }
\right\|_1}>{\left\| {\textbf{x}_{{\textbf{T}^c}}^ * } \right\|_1}.
\end{equation}

Hence, $\textbf{x}^*$ is the optimal solution of (\ref{eq:2}).
Thus, $\textbf{x}^{(1)}=\textbf{x}^*$. The sufficiency is proved.
\end{proof}

\section{Proof of Theorem 2}
\begin{proof}
Suppose $\textbf{Z}$ can be split as $\textbf{Z}=\textbf{Z}_e \cup \textbf{Z}_f$, where $\textbf{Z}_e$ denotes the set composed by the $S_w$ quads that can be recovered, $\textbf{Z}_f = \textbf{Z} \setminus \textbf{Z}_e$. For a quad $\zeta$, we have
\begin{equation}\label{pze}
\textbf{P}(\zeta \in \textbf{Z}_e) =  \frac{S_w}{C_n^\ell  C_{\ell}^{p_2}  C_{n-\ell}^{p_1}  2^{\ell-p_2}}
\end{equation}
Now we define a sequence of random variables $\upsilon_k$ using the set of of quads $\textbf{Z}_e$
\begin{equation}
\upsilon_k=\left\{
             \begin{array}{ll}
               1, & \zeta_k \in \textbf{Z}_e \\
               0, & \zeta_k \in \textbf{Z}_f
             \end{array}
           \right.
\end{equation}
where $k=1,2,..., \zeta_k$ is a quad randomly taken from $\textbf{Z}$.

From (\ref{pze}), it follows that $\textbf{P}(\upsilon_k=1)={S_w}/{C_n^\ell  C_{\ell}^{p_2}  C_{n-\ell}^{p_1}  2^{\ell-p_2}}$, $\textbf{P}(\upsilon_k=0)=1-\textbf{P}(\upsilon_k=1)$. Therefore, $\upsilon_k, k=1,2,...$ are independent and identically distributed random variables with the expected value $E(\upsilon_k)={S_w}/{C_n^\ell  C_{\ell}^{p_2}  C_{n-\ell}^{p_1}  2^{\ell-p_2}}$.

According to the law of large numbers (Bernoulli) in probability theory, the sample average $(1/M)\sum\nolimits_{i = 1}^M {{V_i}}=K/M$ converges towards the expected value $E(\upsilon_k)$, where ${V_i}$ is a sample of the random variable $\upsilon_k$. It follows that when $M$ is sufficiently large
\begin{equation}
E(\upsilon_k)={S_w}/{C_n^\ell  C_{\ell}^{p_2}  C_{n-\ell}^{p_1}  2^{\ell-p_2}} \backsimeq \frac{K}{M}
\end{equation}
The theorem is proven.
\end{proof}

\bibliographystyle{IEEETran}
\bibliography{RAofMCS}

\end{document}